\documentstyle[12pt]{article}

\fnsymbol{footnote}

\newcommand{\be}{\begin{equation}}
\newcommand{\ee}{\end{equation}}
\newcommand{\bea}{\begin{eqnarray}}
\newcommand{\eea}{\end{eqnarray}}

\def\tr{{\mathrm Tr\,}}                        
\def\G{{\cal G}}   
\def\Z{{\mathbf Z}}
\def\M{{\cal M}}
\def\H{{\cal H}}
\def\A{{\cal A}}
\def\D{{\cal D}}
\def\R{{\mathbf R}}
\def\cR{{\cal R}}
\def\pa{{\partial}}

\begin{document}

\thispagestyle{empty}
\setcounter{page}{0}


\begin{center} 
\Large \bf Wakimoto realizations of current and exchange 
algebras\footnote{Talk given at the $7^{th}$ 
Colloquium on Quantum Groups and Integrable Systems, 
Prague, June 1998.} \\
\end{center}

\vspace{.1in}

\begin{center}
L. Feh\'er\\
\vspace{0.3in}
{\em Department of Theoretical Physics, J\'ozsef Attila University \\
Tisza Lajos krt 84-86, H-6720 Szeged, Hungary \\
e-mail: lfeher@sol.cc.u-szeged.hu }
\end{center}

\vspace{.3in}

\begin{center} {\bf Abstract} \end{center}

Working at the level of Poisson brackets, we describe 
the extension of the generalized 
Wakimoto realization of a simple Lie algebra valued current, $J$, 
to a corresponding realization of a group valued chiral primary field, $b$,
that has diagonal monodromy and  satisfies $Kb'=Jb$.
The chiral WZNW field $b$ is subject to a monodromy dependent exchange 
algebra, whose derivation is reviewed, too.
 
\newpage

\section{Introduction}

Affine Lie algebras are symmetries of interesting integrable systems.
In order to perform computations, it is often useful 
to realize the simple Lie algebra valued 
current that generates the symmetry and the associated 
primary fields in terms of free fields.
A remarkable family of generalized
free field realizations of current algebras are the Wakimoto realizations
(see e.g.~the reviews in \cite{BMP,F}).
Recently \cite{dBF}  an explicit formula was found for the Wakimoto
realization of the current in the general case. 
In the context of the WZNW model \cite{W}, it is 
natural to introduce a chiral, group valued
primary field, $b$, which is related to the 
current, $J$, by the differential equation $Kb'=Jb$ and has 
diagonal monodromy.
The Poisson bracket (PB) (also called ``classical exchange algebra'') 
for such a chiral WZNW field 
is given \cite{BDF} in terms of a distinguished solution
of the classical dynamical Yang-Baxter equation.
The main purpose of this paper  is to present an explicit formula
that extends the generalized Wakimoto realization of the current algebra
to a companion realization of the exchange algebra.

\section{Wakimoto realizations of the current algebra}

Let $\G$ denote a simple complex Lie 
algebra\footnote{
All subsequent formulas are valid for  
the normal (split) real form of $\G$, too.}. 
Consider the loop algebra $\widetilde{\G}$ whose elements
are $\G$-valued,
smooth, $2\pi$-periodic functions on the real line $\R$.
Let $\widehat{\G}$ be the standard central extension of $\widetilde{\G}$.
Fixing the value of the dual of the central element to a  constant, $K$, 
one obtains a hyperplane $(\widehat{\G})^*_K$
in the smooth dual $(\widehat{\G})^*$ of $\widehat{\G}$.
The Lie-Poisson bracket  on $(\widehat{\G})^*$ restricts to the usual
current algebra PB on $(\widehat{\G})^*_K$.
Using the model
$(\widehat{\G})^*_K=\{ J\,\vert\, J\in \widetilde{\G} \}$,
the current algebra PB  has the form  
\be
\{ \tr(T_aJ)(x), \tr(T_b J)(y)\}= \tr([T_a, T_b] J)(x)\delta(x-y)
+ K\tr(T_a T_b) \delta'(x-y),
\label{JPB}\ee
where $\delta(x-y)= {1\over 2\pi} \sum_{n\in \Z} e^{in (x-y)}$,
$\{ T_a\}$ is a basis of $\G$ 
and $\tr$ stands for an invariant scalar product on $\G$.
To  describe the generalized Wakimoto realization
of this PB, consider now 
 a triangular decomposition, ${\cal G}={\cal G}_-+{\cal G}_0+{\cal G}_+$, 
defined by means of some integral gradation of $\cal G$.
Denote by $G$ a connected Lie group whose Lie algebra is 
$\G$, and let $G_{0,\pm}\subset G$ be the subgroups 
corresponding to $\G_{0,\pm}\subset \G$.
Then consider the manifolds $\widetilde{G}_+$ and $\widetilde{\G}_-$
whose elements are smooth, $2\pi$-periodic functions on $\R$ with values
in $G_+$ and $\G_-$, respectively.
By means of left translations, identify the cotangent bundle
of $\widetilde{G}_+$ as 
$T^* \widetilde{G}_+ = \widetilde{G}_+ \times \widetilde{\G}_- =
\{ (\eta_+, i_-)\,\vert\, \eta_+\in \widetilde{G}_+,\,\,\, 
i_-\in \widetilde{\G}_-\,\}$,
and equip it  with the canonical symplectic form 
$- d \int_0^{2\pi} \tr ( i_- \eta_+^{-1} d \eta_+)$.
The corresponding PB on  $T^* \widetilde{G}_+$ is encoded by
\begin{eqnarray}
&&\{\,\tr(V^\alpha\, i_-)(x) \,, \, \tr(V^\beta\, i_-)(y) \,\}
     = \tr([V^\alpha,V^\beta]\, i_-)(x) \,\delta(x-y) ,
\nonumber \\
&& \{\,\tr (V^\alpha\, i_-)(x) \,, \, \eta_+(y) \,\}
     =  \eta_+(x)V^\alpha \,\delta(x-y) ,
\quad
\{ \eta_+(x), \eta_+(y)\}=0,
\label{iPB}\end{eqnarray}
where $V^\alpha$ is a basis of $\G_+$.
Now the generalized Wakimoto realization of the current algebra PB 
based on $\G$ is given by a {\em Poisson map} 
$W: (\widehat{\G}_0)^*_K \times T^* \widetilde{G}_+ \rightarrow
\widehat{\G}^*_K$,
where $(\widehat{\G}_0)^*_K =\{ i_0\}$ is the space of $\G_0$-valued
currents endowed with the PB 
\be
\{ \tr(Y_k i_0)(x), \tr(Y_l i_0)(y)\}= \tr([Y_k, Y_l] i_0)(x)\delta(x-y)
+ K\tr(Y_k Y_l) \delta'(x-y)
\label{i0PB}\ee
with a basis $Y_k$ of $\G_0$.
In \cite{dBF} the following simple formula was obtained:
\be
W: (\widehat{\G}_0)^*_K \times T^* \widetilde{G}_+
\ni (i_0, \eta_+, i_- ) \mapsto J=\eta_+(i_0 -i_- ) \eta_+^{-1} + K 
\eta_+' \eta_+^{-1}\in (\widehat{\G})^*_K.
\label{classwak}\ee 
The PBs of $J$ in  (\ref{JPB}) follow from the simpler PBs of the 
constituents $(i_0, \eta_+, i_-)$.

\section{The monodromy dependent exchange algebra} 

In the context of the WZNW model, it is natural to introduce a chiral
$G$-valued field $b(x)$ with the following properties:

\begin{enumerate}
\item The field $b(x)$ is related to the current $J(x)$ by the equation
$J= K b' b^{-1}$.
\item The monodromy matrix associated with $b$ is diagonal:
$b(x+2\pi)= b(x) e^{\omega}$ with $\omega \in \H\subset \G$,
where $\H$ is a fixed Cartan subalgebra of $\G$.
\item
The space of fields $\{ b\}$ has a PB, which is such that 
it implies the current algebra PB for $J=K b' b^{-1}$ and
$b$ is a primary field:
$$
\{ b(x), J_a^n\} = {1\over 2\pi} e^{-inx } T_a b(x) 
\quad\hbox{for}\quad 
J_a^n:={1\over 2\pi} \int_0^{2\pi} dx\, e^{-inx} \tr(T_a J)(x).
$$
\end{enumerate}

A PB satisfying conditions 1--3 is given, for 
$0<\vert x-y\vert < 2\pi$,  by
\begin{eqnarray}
&&\{ b_1(x), b_2(y)\} = {1 \over 2 K} \left(b(x)\otimes b(y)\right)
\Bigl( {\rm{sign}}(y-x) {\cal C} + {\cR}(\omega) \Bigr),
\nonumber\\
&&{\cR}(\omega) =\sum_{\alpha\in \Phi} {\vert \alpha\vert^2 \over 4}
\coth\left({1\over 2} \alpha(\omega)\right)
( E_\alpha \otimes E_{-\alpha} - E_{-\alpha} \otimes E_\alpha).
\label{bbPB}
\end{eqnarray}
Here $\Phi$ denotes the set of roots of $(\H,\G)$,  the
root vectors $E_\alpha$ are normalized by 
$\tr(E_\alpha E_{-\alpha})={2\over \vert \alpha \vert^2}$,
${\cal C}= \sum_a T^a \otimes T_a$ 
is the tensor Casimir  of $\G$ and $b_1=b\otimes {\mathbf 1}$.

In \cite{BDF} 
(see also \cite{BBT}) the ``classical exchange algebra''
(\ref{bbPB})  was obtained by means of a 
construction of the field $b(x)$ out of some
local coordinates on the phase space of the WZNW model. 
As a consistency check, it was verified that ${\cR}(\omega)$
satisfies the following dynamical version of the 
modified classical Yang-Baxter equation:
\be
[\cR_{12}(\omega), \cR_{23}(\omega)] + 2 \sum_i H_1^i  
{\pa \over \pa \omega^i}
\cR_{23}(\omega)
+ \hbox{cycl.~perm.} = 
- \sum_{a,b,c} f^{abc} T_a\otimes T_b \otimes T_c,
\label{CDYB}\ee
where $\{ H^i\} $ is basis of $\H$, $\omega^i = \tr(H^i \omega)$ and
$[T_a,T_b]=\sum_c f_{ab}^{{\phantom{ab}}c}\, T_c$. 

Let us comment on the uniqueness of the PB (\ref{bbPB})
under the conditions  1--3.
It is not difficult to see that  conditions 1--3 imply that
the PB $\{ b_1(x), b_2(y)\}$ must be of the form that
appears in (\ref{bbPB}) with {\em some} r-matrix $\cR(\omega)$.
It also follows that the r-matrix in question must satisfy 
(\ref{CDYB}) and it must be neutral with respect to $\H$.
Recently \cite{EV} all neutral solutions of (\ref{CDYB})  have been
classified  under the additional assumption that  
$\cR: \H\rightarrow \G\wedge\G$ is a meromorphic function.
The parameters that label   
the general solution contain a regular semisimple 
subalgebra of $\G$ and an element of $\H$.
The particular r-matrix in (\ref{bbPB}) corresponds to  
the regular subalgebra being equal to $\G$ and 
the element of $\H$ being zero.
It can be shown \cite{BFP} that 
in the context of the WZNW model no other 
parameters could be chosen, and conditions 1--3 
determine the exchange algebra of the field $b$ essentially uniquely.

The PB in (\ref{bbPB}) can actually be derived from a 
symplectic form \cite{G,Chu,FG} on the phase space 
\be
\M_{G}^{\mathrm Bloch}=\{ b \in C^\infty(\R, G)\,\vert\,
b(x+2\pi)= b(x) e^{ \omega},
\quad 
\omega\in \A \subset \H\,\},
\label{Bloch}\ee
where $\A$ is an open Weyl alcove that 
contains  such $\omega\in \H$
for which $\alpha(\omega)\notin i2\pi \Z$ for any $\alpha\in \Phi$ 
and the restriction of the map $\omega \mapsto e^{ \omega}$ 
to $\A$ is injective.
The symplectic form on $\M_{G}^{\mathrm Bloch}$ is defined by
\be
\Omega_{G,K}^{\mathrm Bloch}(b)= 
- {K\over 2 }\int_0^{2\pi}
 {\tr}\left(b^{-1}d b \right) \wedge
\left(b^{-1}db \right)'
-{K\over 2 }  {\tr}\left((b^{-1} d b)(0)\wedge d \omega\right).
\label{Blochform}\ee 
The condition $\alpha(\omega)\notin i2\pi \Z$ is needed since at the excluded
values of $\omega$ the 2-form in (\ref{Blochform}) would become singular.
The additional restriction of $\omega$ to $\A$ ensures that
$\omega$ uniquely parametrizes the monodromy matrix of 
the ``Bloch wave'' $b$, which  can be represented in the form 
$b(x)=h(x) \exp(x \omega /2\pi)$ with $h\in \widetilde{G}$.
This parametrization of $b$ is useful for deriving the PB in (\ref{bbPB})
by inverting the symplectic form in (\ref{Blochform}).
 
Let us recall \cite{G,Chu,FG}  how the symplectic form 
(\ref{Blochform}) arises in the WZNW model. 
The WZNW model can be described as the Hamiltonian system 
$(\M, \Omega_{K}, H_{\mathrm WZ})$, where
$$
\M= T^* {\widetilde G}= \{\, (g,J)\,\vert\,
g\in \widetilde{G},\,\,\, J\in \widetilde{\G}\,\},
$$ 
$$
\Omega_{K}= d \int_{0}^{2\pi}  {\rm Tr}\left( J d g g^{-1}\right)
+ {K\over 2 }\int_{0}^{2\pi}
 {\rm Tr}\left(d g g^{-1}\right) \wedge \left(d g g^{-1}\right)',
$$ 
and  $H_{\mathrm WZ}={1\over 2K} \int_0^{2\pi} \tr\left(J^2 +I^2\right)$
with $I= -g^{-1} Jg + K g^{-1} g'$. The corresponding 
space of solutions, $\M^{\mathrm sol}$,  
consists of smooth, $G$-valued functions
$g(\sigma, \tau)$ which are $2\pi$-periodic in the space
variable  $\sigma$ and satisfy $\pa_- (\pa_+ g \cdot g^{-1})=0$,
where $\pa_\pm$ are derivations  
with respect to the light cone coordinates $x^\pm = \sigma \pm \tau$.
The most general such function has the factorized form 
$g(\sigma, \tau)= g_L(x^+) g_R^{-1}(x^-)$ with 
a pair $(g_L, g_R)$ of $G$-valued, smooth,
quasiperiodic functions on ${\bf R}$ with equal monodromies:
$g_L(x + 2\pi) = g_L(x)Q$ and 
$g_R(x+2\pi)= g_R (x) Q$ with some $Q\in G$.
It is therefore convenient to introduce the set  
of such pairs, $\widehat\M:= \{ (g_L, g_R) \}$.
By associating
the elements of the solution space with their initial data at $\tau=0$,
one can identify 
$\M^{\mathrm sol}$ with the phase space $\M$, 
and thus  $\Omega_{K}$ yields  a  symplectic form on $\M^{\mathrm sol}$.
Then using the projection 
$\theta: \widehat\M \rightarrow \M^{\mathrm sol}$ given by 
$\theta: (g_L, g_R)\mapsto g=g_L g_R^{-1}$, 
one obtains 
a closed 2-form, $\widehat{\Omega}_{K}$, on $\widehat\M$ by the pull-back.
Explicitly, 
$\widehat{\Omega}_{K}(g_L, g_R)= \Omega^{\mathrm chir}_{K}(g_L) - 
\Omega^{\mathrm chir}_{K}(g_R)$ with 
$$
\Omega_{K}^{\mathrm chir}(g_L)=
- {K\over 2 }\int_0^{2\pi}
 {\rm Tr}\left(g_L^{-1}dg_L \right) \wedge \left(g_L^{-1}dg_L \right)'
-{K\over 2} \tr \left( (g_L^{-1} dg_L)(0)\wedge dQ\cdot Q^{-1}\right).
$$
Of course, $\widehat \Omega_{K}$ is not
a symplectic form on $\widehat{\M}$,
but its restriction  to any local section of the bundle
$\theta: \widehat{\M} \rightarrow \M^{\mathrm sol}$
yields a symplectic form.
As for the chiral WZNW 2-form $\Omega^{\mathrm chir}_{K}$, it is not 
even closed for general monodromy.
However, upon the restriction 
$
g_L(x)=b(x)$  with $b\in \M^{\mathrm Bloch}_G$,
i.e. by imposing the constraint 
$Q=e^{ \omega}$ with $\omega\in \A$,
it becomes the symplectic form in (\ref{Blochform}).
This derivation of $\Omega_{G,K}^{\mathrm Bloch}$ 
can be found in \cite{G,Chu,FG} and these papers
also contain an outline of the derivation of the exchange algebra 
(\ref{bbPB}) from the symplectic form (\ref{Blochform}).
A complete derivation of (\ref{bbPB}) along these lines is given in 
\cite{BFP}.

\section{Wakimoto realizations of the exchange algebra}

We now wish to complete the construction of the following 
commutative diagram:
\be\begin{array}{ccc}
\M_{G_0}^{\mathrm Bloch} \times T^*\widetilde{G}_+ &
\ \stackrel{{\widehat{W} }}{\Longrightarrow}\  
& \M_{G}^{\mathrm Bloch} \\ {} & {} &\\
{\scriptstyle{{\cal D}_0 \times \mathrm{id}}}\,\, 
\downarrow {\phantom{\scriptstyle{{\cal D}_0 \times \mathrm{id}}}}
 & {} &  {\phantom{\scriptstyle{{\cal D}}}} \downarrow 
\,\, {\scriptstyle{{\cal D}}} \\ {} & {} & \\
(\widehat{\G}_0)^*_K \times T^*\widetilde{G}_+ &
\ \stackrel{{W}} \longrightarrow\  & (\widehat{\G})^*_K \\
\end{array}\label{diagram}\ee
The maps designated by simple arrows have already been described 
and are Poisson maps.
In particular, $W$ is the Wakimoto realization of the current algebra
defined in (\ref{classwak}), and $\cal D$ operates according to 
${\cal D}: b \longmapsto J=K b' b^{-1}$.
$\M_{G_0}^{\mathrm Bloch}$ consists of the $G_0$-valued
Bloch waves with regular, diagonal monodromy,
$$
\M_{G_0}^{\mathrm Bloch}=\{ \eta_0 \in C^\infty(\R, G_0)\,\vert\,
\eta_0(x+2\pi)= \eta_0(x) e^{\omega},
\quad 
\omega\in \A \subset \H\,\},
$$
where $\A$ also appears in (\ref{Bloch}), 
and is equipped with the symplectic form
$$
\Omega_{G_0,K}^{\mathrm Bloch}(\eta_0)= 
- {K\over 2 }\int_0^{2\pi}
 {\tr}\left(\eta_0^{-1}d\eta_0 \right) \wedge
\left(\eta_0^{-1}d\eta_0 \right)'
- {K\over 2 }  {\tr}\left((\eta_0^{-1} d\eta_0)(0)\wedge d \omega\right).
$$
The map 
$\D_0$ sends $\eta_0$ to $i_0=K\eta'_0 \eta_0^{-1}\in (\widehat{\G}_0)^*_K$.
The formula for the mapping 
$$
\widehat{W}: \M_{G_0}^{\mathrm Bloch} \times T^*\widetilde{G}_+ \ni
(\eta_0, \eta_+, i_-) \longmapsto b\in \M_{G}^{\mathrm Bloch}
$$
can be found from the equation 
$\D \circ \widehat{W}= W \circ (\D_0 \times {\mathrm id})$, 
which requires that
$$
K b'b^{-1}= \eta_+(K \eta_0'\eta_0^{-1} -i_- ) 
\eta_+^{-1} + K \eta_+' \eta_+^{-1}.
$$ 
A solution for $b$ exists that admits a generalized Gauss decomposition.
In fact,  
$$
b(x)=b_+(x) b_0(x) b_-(x)
\quad\hbox{with}\quad  b_{\pm, 0}(x)\in G_{\pm, 0}
$$
is a solution if
$$
b_+=\eta_+, 
\quad
b_0=\eta_0
\quad
\hbox{and}\quad
K b_-' b_-^{-1}= -  \eta_0^{-1} i_- \eta_0.
$$
The general solution of the differential equation for $b_-$ 
can be written in terms of the particular solution $b_-^P$ defined by  
$b_-^P(0)={\mathbf 1}$ as $b_-(x)= b_-^P(x)S$ with an arbitrary  
$S\in G_-$. 
The constant $S=b_-(0)$ has to be determined from the condition 
that $b$ should have diagonal monodromy.
One finds that $b$ has diagonal monodromy,
indeed it satisfies $b(x+2\pi)= b(x) e^{ \omega}$,
if and only if 
$$
e^{-\omega} S e^{ \omega } = b_-^P(2\pi) S.
$$
Inspecting this equation grade by grade 
using  a parametrization $S=e^s$, $s\in \G_-$ and 
the regularity of $\omega$, one sees that it has 
a unique solution for $S$ as a function of $\omega$ and $b_-^P(2\pi)$. 
Determining $S$ by this procedure, we now define the  mapping
\be
{\widehat W}: 
\M_{G_0}^{\mathrm Bloch} \times T^*\widetilde{G}_+ \ni
(\eta_0, \eta_+, i_-) \longmapsto b=\eta_+ \eta_0 b_-^P S 
\in \M_{G}^{\mathrm Bloch}
\label{wakiext}\ee
that makes the diagram in (\ref{diagram}) commutative.

\medskip
\noindent
{\bf Theorem.}
{\em The mapping $\widehat{W}$ defined in (\ref{wakiext})
is symplectic, that is, }
\begin{eqnarray}
&&(\widehat{W}^* \Omega_{G, K}^{\mathrm Bloch})(\eta_0, \eta_+, i_-)=
\Omega_{G,K}^{\mathrm Bloch}(b=\eta_+ \eta_0 b_-^P S)=\nonumber \\
&&\qquad\qquad \qquad 
\Omega_{K, G_0}^{\mathrm Bloch}(\eta_0) 
- d \int_0^{2\pi} \tr ( i_- \eta_+^{-1} d \eta_+).
\nonumber\end{eqnarray}

\medskip
The second term on the right hand side is the 
symplectic form on $T^*\widetilde{G}_+$.
The proof of the theorem \cite{BFP} is a straightforward computation.
For $G=SL(2)$ the result was already proved in 
\cite{AS,G,FG} and some other
special cases of it can be extracted from \cite{GMOMS}.

Since a symplectic map is always  a Poisson map too,
$\widehat{W}$ provides us with a realization of the monodromy 
dependent exchange algebra (\ref{bbPB}) of the $G$-valued Bloch waves 
in terms
of the analogous exchange algebra of the $G_0$-valued Bloch waves and 
the canonical free fields that may be used to parametrize
$T^*\widetilde{G}_+$.
This generalized free field realization becomes a true free field
realization in the principal
case, for which $\G_0=\H$ is Abelian and $\eta_0$ is the exponential
of a $\H$-valued free scalar field.

The Wakimoto realizations of the affine Lie algebras in 
generalized Fock spaces, i.e.~at the level 
of vertex algebras  as opposed to the above
Poisson algebras, have many applications in conformal field theory
(see e.g.~\cite{BMP,F}). 
A simple explicit formula for such realizations
of the current $J$  is obtained in \cite{dBF} by 
quantizing the expression in (\ref{classwak}).
It would be interesting to also quantize the formula in (\ref{wakiext}).
Another problem for future study is to work out an analogue of the 
construction presented in this paper for 
the case of $q$-deformed affine Lie algebras.

\bigskip 

\noindent
{\bf  Acknowledgments.}
I wish to thank J.~Balog and L.~Palla for many discussions
on the subject of this report.
This work was supported 
by the Hungarian National Science Fund (OTKA) under T025120.

\end{document}